\documentclass{edm_article}
\usepackage[table]{xcolor}
\usepackage{url}
\usepackage{hyperref}
\usepackage{graphicx}
\usepackage{tabularx}
\usepackage[linesnumbered,ruled,vlined]{algorithm2e}
\usepackage{multirow}
\usepackage{siunitx}

\begin{document}

\title{Starting Seatwork Earlier as a Valid Measure of\newline Student Engagement}

\numberofauthors{7}
\author{
\alignauthor
Ashish Gurung\\
       \affaddr{Carnegie Mellon University}\\
       \email{agurung@cmu.edu}
\alignauthor
Jionghao Lin\\
       \affaddr{University of Hong Kong}\\
       \email{jionghao@hku.hk}
\alignauthor
Zhongtian Huang\\
       \affaddr{University of Pennsylvania}\\
       \email{skyrover@upenn.edu}
\and
\alignauthor
Conrad Borchers\\
       \affaddr{Carnegie Mellon University}\\
       \email{cborchers@cmu.edu}
\alignauthor
Ryan S. Baker\\
       \affaddr{University of Pennsylvania}\\
       \email{ryanshaunbaker@gmail.com}
\alignauthor
Vincent Aleven\\
        \affaddr{Carnegie Mellon University}\\
       \email{aleven@cs.cmu.edu}
\and
\alignauthor
Kenneth R. Koedinger\\
        \affaddr{Carnegie Mellon University}\\
       \email{koedinger@cmu.edu}
}

\maketitle

\begin{abstract}

Prior work has developed a range of automated measures (``detectors'') of student self-regulation and engagement from student log data. These measures have been successfully used to make discoveries about student learning. Here, we extend this line of research to an underexplored aspect of self-regulation: students' decisions about when to start and stop working on learning software during classwork. In the first of two analyses, we build on prior work on session-level measures (e.g., delayed start, early stop) to evaluate their reliability and predictive validity. We compute these measures from year-long log data from Cognitive Tutor for students in grades 8–12 (N = 222). Our findings show that these measures exhibit moderate to high month-to-month reliability (G > .75), comparable to or exceeding gaming-the-system behavior. Additionally, they enhance the prediction of final math scores beyond prior knowledge and gaming-the-system behaviors. The improvement in learning outcome predictions beyond time-on-task suggests they capture a broader motivational state tied to overall learning. The second analysis demonstrates the cross-system generalizability of these measures in i-Ready, where they predict state test scores for grade 7 students (N = 818). By leveraging log data, we introduce system-general naturally embedded measures that complement motivational surveys without extra instrumentation or disruption of instruction time. Our findings demonstrate the potential of session-level logs to mine valid and generalizable measures with broad applications in the predictive modeling of learning outcomes and analysis of learner self-regulation. 

\end{abstract}

\section{Introduction}
\label{section:intor}
Self-regulated learning (SRL)~\cite{pintrich2004conceptual} has been a cornerstone in the development of automated detectors of student engagement in Educational Data Mining (EDM). Given the nature of logged data from Digital Learning Platforms (DLP), EDM researchers have frequently leveraged time-based measures to capture key aspects of self-regulation, such as motivation, time management, and goal orientation. These time-based measures have been used to contextualize various aspects of learner behaviors in the development of detectors for effort~\cite{shih2011response,gurung2021examining}, attention~\cite{yonetani2012multi,pham2015attentivelearner}, gaming-the-system~\cite{paquette2014towards}, mind-wandering~\cite{christoff2018mind}, and off-task behavior~\cite{smallwood2015science,baker2004off}. Follow-up research has demonstrated the predictive utility of these measures, linking them to students' immediate~\cite{cocea2009impact,smallwood2015science,gurung2021examining}, short-term, and long-term learning outcomes~\cite{pardos2013affective,san2014predicting}.   

While EDM and Learning Analytics have extensively examined process-level engagement\textemdash focusing on interactions with tasks or materials\textemdash session-level exploration has been generally limited to students working independently in higher education\cite{gasevic2017detecting,fan2021learning,liu2015modeling,kizilcec2017self}. Despite an inherent expectation for student participation in K-12 settings, session-level engagement remains largely underexplored. Beyond a few studies on session-level time-based measures as indicators of diligent learner behavior~\cite{dang2017detecting,dang2020ebb}, its broader implications are not well understood. As such, dedicated classwork sessions provide a valuable opportunity to investigate session-level behaviors in structured learning environments.

Student behavior during classwork sessions can be influenced by individual self-regulation~\cite{zimmerman2002becoming,gollwitzer2006implementation} as well as broader classroom dynamics, and teacher support~\cite{furrer2014influence,pintrich2004conceptual}. For instance, individual goal orientation~\cite{gollwitzer2006implementation,midgley2001performance} as well as teacher instructions, peer engagement, or other systemic factors~\cite{mctighe2005seven,brookfield2015skillful,pintrich2004conceptual} can influence a student's decision about when to start. 

Addressing this gap, especially for classwork sessions, is essential for understanding how classroom dynamics shape individual learning behaviors and overall learning outcomes. Additionally, combining session-level and process-level engagement can provide deeper insights into how behaviors at both levels influence broader engagement and learning.

In this light, we build on prior research leveraging session-level time-based measures~\cite{dang2017detecting,yang2021exploring} to examine their reliability and predictive validity in relation to learning outcomes, aiming to identify student differences in diligent time use. Specifically, we analyze how students' behavior, such as delayed starting, session length, and early stopping when using DLP during classwork, manifests across students and influences their learning outcomes, including year-end grades and performance on standardized state tests. Through these session-level measures, we contribute to the growing body of EDM and Learning Analytics research mining SRL differences from log data, offering a novel lens on student-initiated practice behavior in the context of a classwork session. 


\section{Background}
\label{section:pw_bg}
\subsection{Self-Regulated Learning}
The conceptualization of SRL as a theoretical construct has evolved over time, with several empirical studies contributing to its refinement~\cite{panadero2017review}. Among the formalizations of SRL, the models proposed by Zimmerman~\cite{zimmerman2002becoming}, Pintrich~\cite{pintrich2004conceptual}, and Winne and Hadwin~\cite{WinneHadwin1998} are often referenced in EDM and Learning Analytics. These frameworks offer distinct yet complementary perspectives on factors and mechanisms influencing learner behavior. Zimmerman's model emphasizes individual self-regulation, whereas Pintrich's framework, as well as Winne and Hadwin's framework, emphasize the importance of contextual influences (e.g., peer interactions, teacher guidance, and class norms), making it more suitable for our focus on K-12 classwork session data.


Previous research has explored the influence of various SRL factors on self-regulatory processes. Goal setting, task value, and goal orientation influence engagement, with mastery goals fostering deep learning~\cite{dweck1986motivational} and performance goals prioritizing learning outcomes~\cite{midgley2001performance,gollwitzer2006implementation}. Motivation and self-efficacy underpin all phases, shaping persistence and self-regulation~\cite{wolters2003regulation,eccles2020expectancy}, while high self-efficacy fosters resilience, helping learners reframe challenges as opportunities~\cite{winne2017cognition}. Additionally, time management, effort regulation, and metacognitive monitoring sustain engagement by guiding progress tracking, strategic adjustments, and reflection~\cite{winne2011cognitive,azevedo2013metacognition}.

In practice, effective self-regulation is often observed through learners' adjustments in how they manage time and effort. For example, Gollwitzer et al.~\cite{gollwitzer2006implementation}, in their meta-analysis, highlighted how goal-oriented, explicit ``if-then'' plans shape student self-regulation. For instance, a student might plan, \textit{``If my friends distract me, then I will ignore them and keep working.''} These situational cues helped students effectively regulate their learning by developing strategic goal-directed plans for seizing opportunities(\textit{d} = 0.61) and shielding them from unwanted influences (\textit{d} = 0.77). 

\subsection{Time-Based Detectors}
Different dimensions of SRL have directly or indirectly informed the development of time-based measures (i.e., detectors) to capture various aspects of learner behavior in DLPs. These measures provide valuable insight into various aspects of student engagement, such as effort~\cite{shih2011response,gurung2021examining}, off-task behavior~\cite{smallwood2015science,baker2004off}, and procrastination~\cite{borchers2025workload,yao2020analyzing,sabnis2022large}.

Learner time management also plays a crucial role in the modeling and detection of more complex behaviors, such as gaming-the-system~\cite{baker2008developing,paquette2014towards}, wheel-spinning (i.e., unproductive persistence)~\cite{beck2013wheel,gong2015towards}, attention~\cite{yonetani2012multi,pham2015attentivelearner}, and mind-wandering~\cite{christoff2018mind,mills2016automatic}. For instance, the rule-based gaming-the-system detector by Paquette et al.~\cite{paquette2014towards} uses 19 contextually interpretable actions, 13 of which ($\sim$68\%) directly incorporate time to establish context. For example, a help request within five seconds of starting a problem would indicate: \textit{``the student did not think before asking for help.''}

Follow-up research has reported on the predictive performance of some of these behaviors in relation to students’ immediate~\cite{cocea2009impact,smallwood2015science,gurung2021examining}, short-term, and long-term learning outcomes~\cite{pardos2013affective,san2014predicting}. For instance, in terms of immediate outcomes, Gurung et al.~\cite{gurung2021examining} found that students who exhibit low effort (time spent) on hints were significantly more likely to wheel-spin on mastery-based assignments. Similarly, for long-term outcomes, San Pedro et al.~\cite{san2014predicting} observed a significant relationship between students' frequency of gaming-the-system while doing math and their eventual choice of STEM versus non-STEM careers, i.e., students pursuing non-STEM careers exhibited gaming-the-system behavior 0.57 standard deviations more frequently than their peers.

Despite advances in process-level detectors, few studies have incorporated session-level measures (e.g., start time, session length, stop time) to capture student engagement in structured learning environments. Factors such as goal orientation~\cite{midgley2001performance,gollwitzer2006implementation}, procrastination~\cite{borchers2024you,yao2020analyzing}, and teacher/peer influence~\cite{brookfield2015skillful,pintrich2004conceptual} can affect when students start their work. Similarly, metacognitive monitoring~\cite{azevedo2013metacognition}, managing distractions~\cite{gollwitzer2006implementation}, self-efficacy~\cite{wolters2003regulation,eccles2020expectancy}, and other contextual factors~\cite{brookfield2015skillful,pintrich2004conceptual} can influence how long they continue before stopping. Examining these session-level measures, both individually and relative to peers, furthers our understanding of how students self-regulate to balance intrinsic and extrinsic influences, ultimately facilitating more diligent time use.

\subsection{Diligent Use of Time}
Diligence is characterized by sustained effort and perseverance in learning~\cite{bernard1993diligence}. While some research has examined diligence as a distinct construct~\cite{bernard1993diligence,galla2014academic}, it is often conceptualized as a behavioral manifestation rather than an independent trait. This perspective frequently links diligence to broader frameworks such as behavioral regulation in SRL~\cite{zimmerman1990self}, conscientiousness~\cite{kim2016conscientiousness}, or grit~\cite{duckworth2007grit}. While diligence has not been as widely studied as SRL, aspects of the construct intuitively align with teacher expectations and interpretations of student engagement patterns~\cite{brookfield2015skillful} in terms of factors such as persistence, time management, and punctuality. 

Dang et al.\cite{dang2017detecting} extended the Academic Diligence Task (ADT)\footnote{Galla et al.\cite{galla2014academic} used a computer-based measure, where students choose between practicing math problems or engaging in a more entertaining distractor to evaluate diligence.} proposed by Galla et al.\cite{galla2014academic} to examine learner diligence using log data from a DLP. 
They estimate student diligence as a latent construct by analyzing online behaviors within a DLP. They controlled for total time on task, work rate (problems per time), and prior knowledge, using the number of lessons mastered as the outcome. The inferred diligence demonstrated both predictive and construct validity. For predictive validity, inferred diligence was significantly predictive of final academic scores. Similarly, the measure was correlated with math interest, self-efficacy, mastery approach, and effort regulation. While Dang’s findings underscore the importance of \textit{time on task} in inferring diligence, additional work is required to isolate aspects of student engagement and behavior that influence \textit{time on task}.

In our work, we decompose \textit{time on task} into sub-components such as \textit{delayed start, session length, and early stop}. By focusing on aspects of diligent time use, we aim to align our approach with how teachers assess diligence in their day-to-day practice to set clear expectations, provide timely feedback, and model perseverance~\cite{brookfield2015skillful} at multiple levels, i.e., individually, within peer groups, and across the entire class~\cite{mctighe2005seven}. 

\section{Current Study}
As outlined in Sections~\ref{section:intor} and~\ref{section:pw_bg}, we integrate insights from SRL, diligence, and detectors to develop session-level time-based measures during classwork. Using feature engineering strategies, we capture students’ diligent time use in ways that align with classroom norms and teacher expectations. 

Specifically, we examine the reliability, predictive validity, and generalizability of session-level measures through the following research questions:

\begin{enumerate}
\renewcommand{\labelenumi}{\textbf{RQ\theenumi.}}
\item Are session-level time-based measures of practice stable within students over time? 
\item Are session-level measures of student practice predictive of student learning outcomes?
\item Do these session-level measures of student practice generalize across different educational technologies and classrooms?
\end{enumerate}

\section{Methods}
\label{sec:methodology}

\subsection{Dataset}
\label{sec:Dataset}
In our study, we use two datasets. Dataset 1 is an extended version of the dataset used by Dang et al.~\cite{dang2017detecting}, allowing us to build on their findings and contrast session level measures with \textit{time on task} (\textbf{RQ1} and \textbf{RQ2}). We then use dataset 2 to build on our findings from \textbf{RQ1} and \textbf{RQ2} and examine the generalizability of session-level measures on data from a different learning platform (\textbf{RQ3}).

Dataset 1 is a publicly available\footnote{\scriptsize{\href{https://pslcdatashop.web.cmu.edu/DatasetInfo?datasetId=613}{https://pslcdatashop.web.cmu.edu/DatasetInfo?datasetId=613}}} dataset originally collected over the course of a school year from two suburban middle schools in a mid-Atlantic state in the U.S.\cite{bernacki2013datasource}. The dataset contains student log data from Cognitive Tutor\cite{ritter2007cognitive}, an adaptive math tutoring system. During the school year 2011-2012, 222 students across 12 classes from 2 schools in grades 8 to 12 used Cognitive Tutor for math practice. The average class had 18.5 students (standard deviation (SD) = 3.37). Collectively, these students worked on 23,111 problems, generating more than 2 million transaction logs. The transaction logs provide detailed records of student actions, including correct and incorrect attempts, hint requests, timestamps, problem steps, and associated skills. Additionally, the dataset includes teacher-assigned year-end math grades for current and previous school years.

We use Dataset 2 to examine the generalizability of session-level measures. This dataset contains i-Ready\footnote{\scriptsize{\href{https://i-readycentral.com/articles/middle-school-3/}{https://i-readycentral.com/articles/middle-school-3/}}} log data from a West Coast state in the U.S. i-Ready is an adaptive math tutoring system that provides immediate correctness feedback and help on demand. The dataset contains engagement data from 818 students in grade 7 who used i-Ready for math practice from school years 2022-2023 and 2023-2024. The average class size across 26 classes was 31.06 students (SD = 3.09). Collectively, students worked on 18,909 problems, generating approximately 3.5 million transaction logs. Like Cognitive Tutor, i-Ready data also contains detailed records of student actions, including correct and incorrect attempts, hint requests, timestamps, problem ID, and associated skills. Additionally, the dataset includes students' standardized state test scores for grades 6 (pre) and 7 (post). As the school is in a low-opportunity zone, there was considerable student movement, resulting in an average dropout rate of 4.04 students (SD = 2.78) per class. As such, out of 818 students, only 717 took the state test.

Both datasets were collected in compliance with IRB protocols, ensuring ethical data handling and participant privacy. Parents or guardians had the option to opt their child out of research participation.

\subsection{Identifying Sessions}
Most DLPs do not distinguish between classwork, homework, or additional in-school work (e.g., detention and lunch breaks). Therefore, class session references for aggregating features in this study use automatically inferred metadata for when classes start and end. This section outlines our approach to identifying classwork, homework, and non-classwork sessions from log data.

\subsubsection{Bell Schedules}
A bell schedule is a school's structured timetable that sets start and end times for classes, breaks, and transitions. Bell schedules can provide useful class session metadata; however, relying solely on them presents logistical challenges. This data is often missing from DLP and can be difficult to track. Moreover, DLP usage can deviate from fixed schedules due to administrative tasks (e.g., roll calls and announcements) or pedagogical choices (e.g., structuring class time to combine direct instruction and DLP use). Additionally, students may continue working beyond the scheduled class time, either due to teacher-led extensions or independent motivation. These variations make fixed bell schedules unreliable indicators of session-level metadata.

\subsubsection{Classwork Session}
\label{subsubsec:student_class_session}
We define classwork sessions along two dimensions: (1) the total number of active students and (2) the time threshold (intervals of inactivity), i.e., if the entire class remains inactive for more than a specific interval, we classify the subsequent activity as part of a new session.

For dimension (1), the total number of active students, we draw from prior research in collaborative learning. Students, when collaborating, often work in pairs~\cite{teasley2013constructing} or small groups (3-5)~\cite{stahl2006group} for short periods, whereas classwork typically involves larger groups and longer intervals. As such, we set a threshold of at least five active students to distinguish classwork sessions from sporadic individual or subgroup usage.


For dimension (2), the time threshold, we use Algorithm~\ref{algo:class_inference_algo} to segment student transactions into class sessions based on temporal activity patterns. To ensure the robustness of our approach, we performed a sensitivity analysis, testing threshold values between 2 and 30 minutes to determine the optimal cutoff for inferring sessions.


\begin{algorithm}[htp]
\caption{Class Session Inference Algorithm}
\begin{enumerate}
    \item \textbf{Step 1:} For the class, initialize session and session count.
    \item \textbf{Step 2:} For each transaction: \\
    \textbf{if} time difference between current and last transaction $<$ threshold (e.g., 7.5 minutes)\\
    \textbf{then} \\
        \quad Assign the transaction to the current session. \\
        \textbf{else} \\
        \quad Start a new session. \\
        \quad Increment the session count.
    \item \textbf{Step 3:} When switching to a new class, reset the session count.
\end{enumerate}
\label{algo:class_inference_algo}
\end{algorithm}

\begin{figure}[!ht]
  \Description{There was no major difference in the total inferred sessions. However, higher thresholds ($\geq10 minutes$) resulted in longer session lengths for outliers. As longer sessions can impact the estimation of start time, stop time, and student session length, we used 7.5 minutes as the threshold.}
  \centering
  \includegraphics[width=.905\columnwidth]{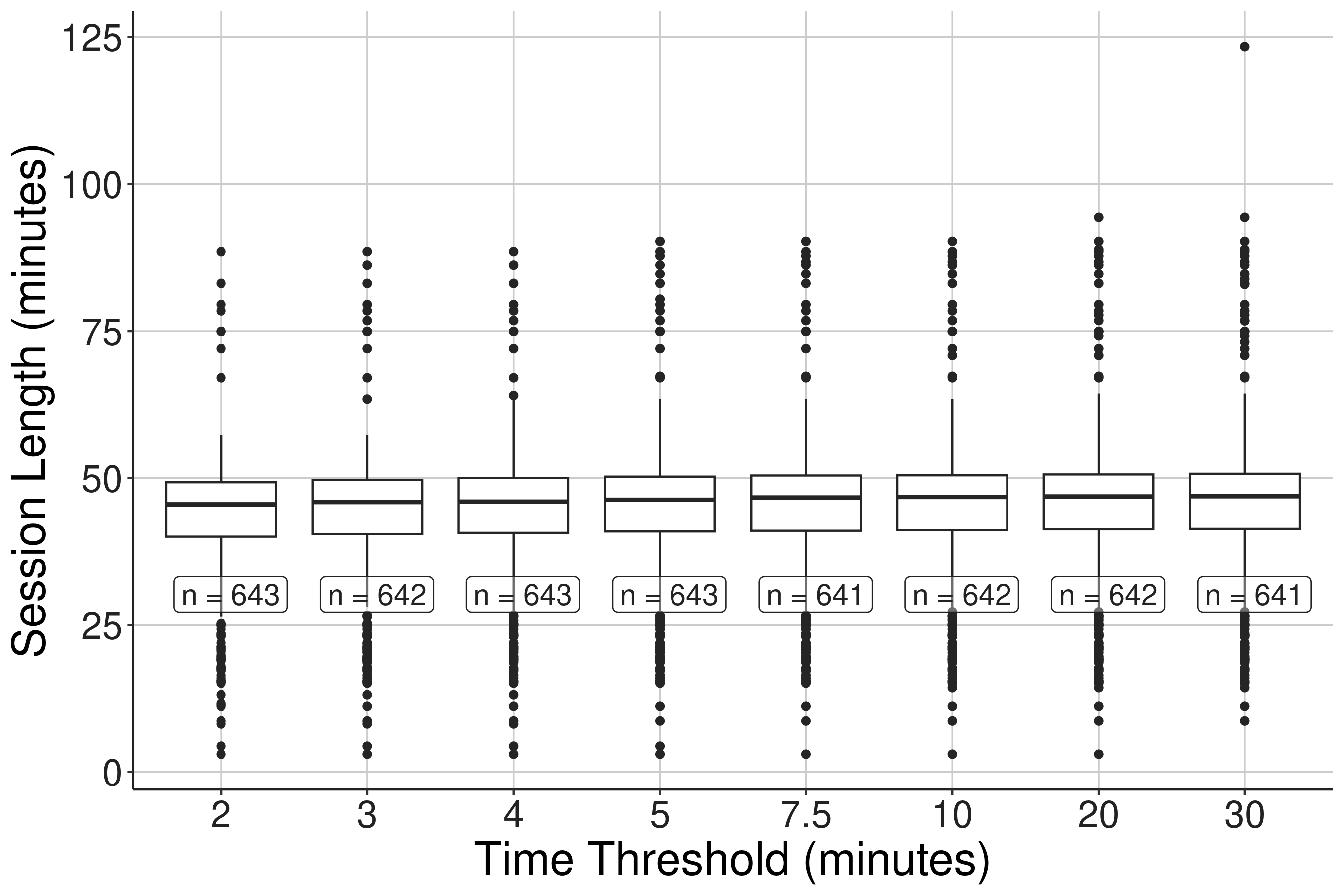}
  \caption{There was no major difference in the total inferred sessions (n). However, higher thresholds ($\geq10\;minutes$) resulted in longer session lengths for outliers. As longer sessions can impact the estimation of \textit{start time, stop time}, and student \textit{session length}, we used 7.5 minutes as the threshold.}
  \label{fig:mathia-threshold-time-session}
\end{figure}

The inferred class sessions with more than five active students using different thresholds are presented in Figure~\ref{fig:mathia-threshold-time-session}. There was no major difference in the total number of inferred sessions. Still, higher thresholds resulted in longer session lengths for outliers, likely due to the inclusion of instances where students logged in during recess or later in the day into the inferred session. Although these occurrences were infrequent, they could still impact the calculation of time-related measures. Given the minimal variation in inferred sessions and the use of 2 to 5 minutes as incremental thresholds for idleness in prior works~\cite{holstein2018student,holstein2019co}, we select 7.5 minutes as the threshold to infer classwork sessions.

The total number of classwork sessions inferred using the 7.5-minute threshold is presented in Table~\ref{tab:descriptive_statistics_classwork}. We also computed the size and length of non-classwork and homework sessions. Notably, the median session size for these sessions was 1, indicating that students were typically by themselves.

\begin{table}[h]
\centering
\caption{Session level information detailing the size and length of classwork, homework, and non-classwork (during school hours) sessions.}
\label{tab:descriptive_statistics_classwork}
\renewcommand{\arraystretch}{1}
\resizebox{\columnwidth}{!}{%
\begin{tabular}{lccccccc}
\hline
                                                 &                         & \multicolumn{2}{c}{{Session Size}}                     & \multicolumn{2}{c}{{Session Length}}                  \\
                                                 &                         & \multicolumn{2}{c}{{(N Students)}}                     & \multicolumn{2}{c}{{(mins)}}                          \\   \cline{3-4} \cline{5-6}
\multicolumn{1}{c}{\multirow{-2}{*}{{Category}}} & {\multirow{-2}{*}{{N}}} & \multicolumn{1}{c}{Mean} &  \multicolumn{1}{c}{Median} & \multicolumn{1}{c}{Mean} & \multicolumn{1}{c}{Median} \\ \hline
 Non Classwork                                       &  485                    &     1.50                 &        1                    &      14.85               &        13.88               \\
 \textbf{Classwork}                              &  \textbf{641}           &     \textbf{14.63}       &        \textbf{15}          &      \textbf{44.99}      &        \textbf{46.65}      \\ 
 Homework                                        &  40                     &     1.35                 &        1                    &      19.18               &        5.43                 \\ \hline 
 \\
\end{tabular}}
\end{table}

\subsubsection{Time-Based Measures}
Guided by prior research on session-level information in K-12 settings~\cite{yang2021exploring,dang2017detecting,dang2022disseration} and insights from broader research on SRL and diligence (Section~\ref{section:pw_bg}), we extend prior works by Dang et al.~\cite{dang2017detecting} to analyze student engagement during classwork sessions. Specifically, we conduct feature engineering of time-based measures that align with signals teachers commonly use to identify diligent time use among students.

\begin{figure*}[!ht]
  \Description{A problem-level visualization of student engagement during classwork using transaction logs. (a) The class session time represents the overall duration of the class, while (b) student session time indicates Student 1's session length. In contrast, Student 2 was absent from this session (c). Additionally, we observe that Student 3 exhibited a (d) delayed start, taking longer to begin work, as well as (e) stopping earlier than their peers. The class session spans from 7:32 AM to 8:25 AM, with a total of 15 students participating.}
  \centering
  \includegraphics[width=.94\linewidth]{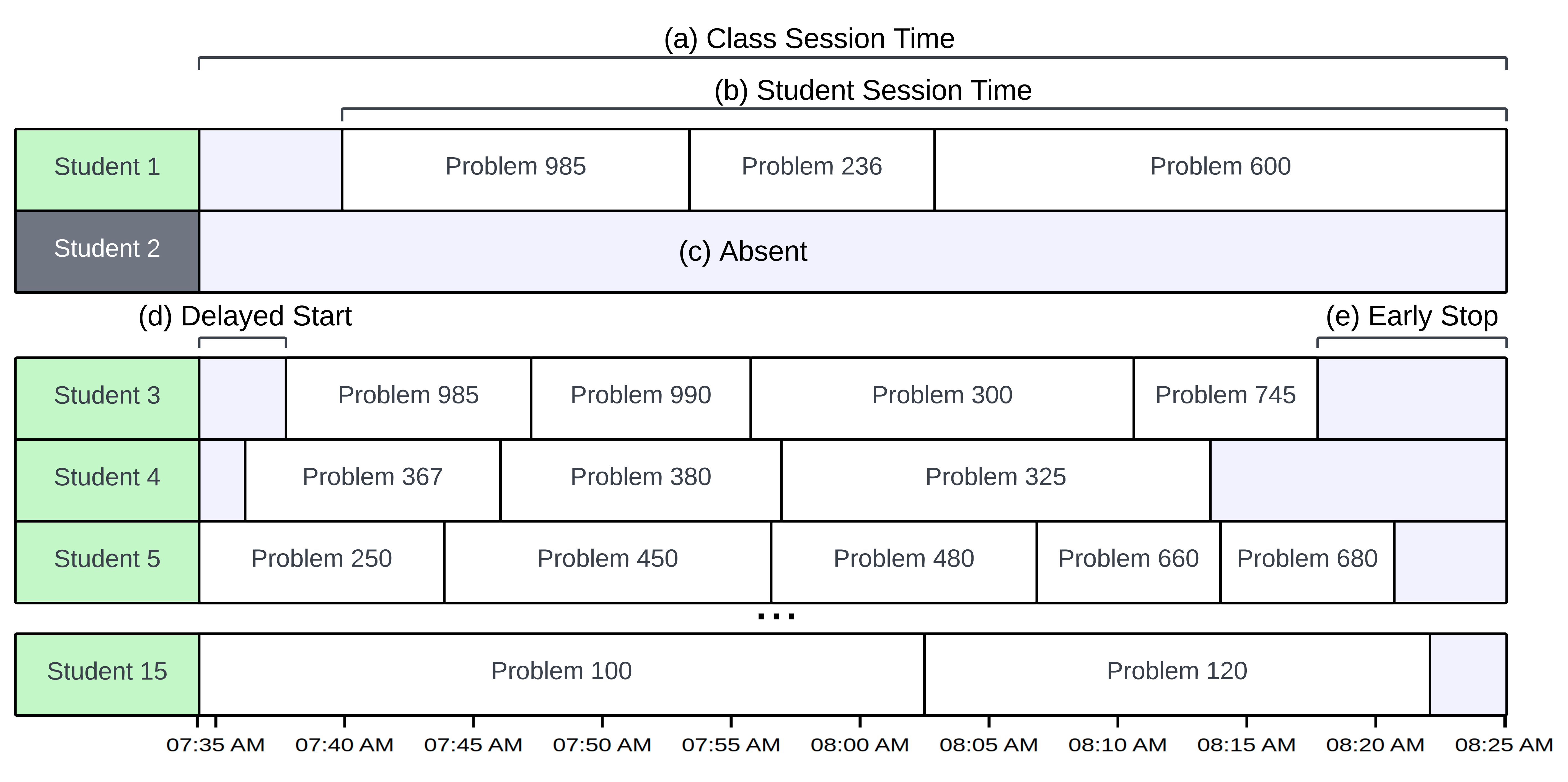}
  \caption{A problem-level visualization of student engagement during classwork. The (a) class session time represents the overall duration of the class, while (b) student session time highlights Student 1's work in Cognitive Tutor. In contrast, Student 2 is (c) absent from this session. Additionally, we observe that Student 3 exhibited a (d) delayed start, taking longer to begin working, and (e) stopped earlier than their peers. The class session spans from 7:32 AM to 8:25 AM and has 15 students.}
  \label{fig:class-session}
\end{figure*}

Figure~\ref{fig:class-session} illustrates a simplified\footnote{Figure~\ref{fig:class-session} is a simplified problem-level view. Each problem contains several transactions to detail the steps taken, the correctness of attempts, and other additional information.} view of a classwork session that informed our feature engineering and identification of potential session-level signals of student engagement. Figure~\ref{fig:class-session} (a) indicates the classwork session occurred from 7:32 AM to 8:25 AM. Figure~\ref{fig:class-session} (b) illustrates the student session length (active time) for Student 1. Figure~\ref{fig:class-session} (c) tells us that Student 2 was absent. Additionally, the visualization also illustrates the delayed start (Figure~\ref{fig:class-session} (d)) and early stop (Figure~\ref{fig:class-session} (e)) behavior of Student 3. The final list of measures is listed in Table~\ref{tab:indicators_learner_behavior}.

\begin{table*}[!ht]
\centering
\scriptsize
\caption{ The list of time-based measures explored in the study.}
\label{tab:indicators_learner_behavior}
\renewcommand{\arraystretch}{1.15}
\resizebox{0.98\textwidth}{!}{%
\begin{tabular}{lp{10.5cm}}
\hline
\textbf{Measure} & \textbf{Description}                                  \\ \hline
delayed start                          & delayed start time during classwork                             \\ 
relative delayed start                 & delayed start time relative to classmates                   \\ 
session time                           & session length during classwork.                                  \\ 
relative session time                  & session length relative to classmates                        \\ 
early stop                             & early stop time during classwork                        \\ 
relative early stop                    & early stop time relative to classmates                \\ 
idle time ($>2mins$)                   & student idle time during classwork                        \\ 
relative idle time ($>2mins$)          & student idle time relative to classmates                \\ 
total time on task                     & total student session time                                       \\ 
usage time ratio                       & ratio of total student session time to total classwork session time              \\ 
relative usage time ratio              & ratio of total student session time to total classwork session time relative to classmates              \\ 
attendance                             & attendance across classwork sessions                                \\ 
\hline
\end{tabular}}
\end{table*}

The \textbf{primary set} of session-level measures we developed was \textit{delayed start, session length, early stop}, and their relative measures. The raw measures help us understand the general patterns in learner behavior, whereas the relative measure helps us understand their behavior with respect to their peers. \textit{Delayed start} and \textit{early stop}, along with their peer-normalized relative measures (\textit{relative delayed start} and \textit{relative early stop}), provide insights into students' ability to monitor and adjust engagement, reflecting aspects of time-management, motivation and effort regulation~\cite{winne2011cognitive,pintrich2004conceptual}. \textit{Session time} and \textit{relative session time} emphasize sustained focus and persistence during learning tasks, often linked to self-efficacy and task persistence~\cite{schunk2012learning}.

Additionally, we also incorporated more well-established measures in \textit{total time on task, attendance} and \textit{student's time use ratio} to capture aspects of time management and self-efficacy~\cite{pintrich2004conceptual}. \textit{Student's time use ratio} is the ratio of a student's total time on task to the total classwork session time for the class. If a student was highly active during and outside of classwork sessions, then this ratio can be greater than 1. Reflecting the student's additional effort and motivation.

Finally, \textit{idle time} and its relative measure help identify periods of inactivity, offering insights into students' ability to maintain focus and manage distractions~\cite{winne2011cognitive,azevedo2013metacognition}. As 99\% of transaction duration was less than $\sim 1.5\; minutes$, we considered student disengagement for $>2\; minutes$ as idle behavior.

By examining these measures during classwork sessions over a school year, we aim to develop an understanding of the students' long-term diligent time use. 

\subsubsection{Gaming Tendency}
Beyond the time-based measures, we also include the students' gaming tendency, as gaming-the-system has been extensively studied in EDM and is recognized to be predictive of both short-term and long-term learning outcomes~\cite{cocea2009impact,pardos2013affective,san2014predicting}. We evaluate the predictive performance of our time-based measures by comparing them with the students' tendency to engage in gaming-the-system behavior. We apply the rule-based gaming-the-system detector (see~\cite{paquette2014towards}) to student transaction logs to identify instances of gaming. 

We then estimate gaming tendencies using a latent modeling approach described by~\cite{dang2022disseration}. This approach is based on a latent modeling approach inspired by item response theory (IRT) that treats gaming tendencies as a stable, trait-like property while accounting for contextual factors, such as task formats and prior knowledge~\cite{huang2023using}. The original model controlled for process-level variance using problem-type information; however, due to the unavailability of problem-type data in our dataset, we used a variation of the model described by~\cite{dang2022disseration}. The modified regression model used for estimating gaming tendencies is detailed in Equation~\ref{eqn:latent_estimation_model}, where gaming tendencies are estimated at the student level ($\gamma_{ij}^{(s)}$) while accounting for class effects ($\gamma_{i}^{(c)}$), and controlling for problem level differences ($\gamma_{k}^{(p)}$).  

\begin{align}
\label{eqn:latent_estimation_model}
\text{logit}({P}(\text{gaming} = 1)) &= \beta_0 + \gamma_i^{(c)} + \gamma_{ij}^{(s)} + \gamma_{k}^{(p)} + \epsilon_{ijk}
\end{align}

\subsection{Analysis Plan}
\label{subsection:analysis-plan}
We employ the following approach to explore our research questions to evaluate the stability, predictive validity, and cross-system generalizability of session-level time-based measures. The preprocessing, identification of sessions and implementation of the rule-based gaming detector were conducted in \textit{Python}. The exploratory and statistical analyses were performed in \textit{R}, using \textit{lme4} and \textit{gtheory} packages.

\subsubsection{Stability of Measures (RQ1):} 
We assessed the stability of time-based measures using Generalizability Theory (G-Theory)\cite{cronbach1972dependability,brennan1992generalizability,brennan2010generalizability}, by evaluating measurement reliability over time. We specified a variance component model for each measure using Equation~\ref{eqn:gtheory_eqn}. The model uses random intercept to account for individual differences $\gamma_j^{(student)}$ while capturing variability across months $\gamma_l^{(month)}$ for each observation $Y_{jl}$. We computed G-coefficients (Equation\ref{eqn:g-coefficient}) to assess the overall reliability, also called generalizability, of each measure and $\phi$-coefficients (Equation~\ref{eqn:phi-coefficient}) to determine their consistency for decision-making applications.

\begin{align}
\label{eqn:gtheory_eqn}
 Y_{jl} &= \beta_0 + \gamma_j^{(student)} + \gamma_{l}^{(month)} + \epsilon_{jl}
\end{align}

\begin{equation}
\label{eqn:g-coefficient}
G = \frac{\sigma^2_{student}}{\sigma^2_{student} + \sigma^2_{residual}}    
\end{equation}

\begin{equation}
\label{eqn:phi-coefficient}
\phi = \frac{\sigma^2_{student}}{\sigma^2_{student} + \sigma^2_{residual} + |\sigma^2_{month}|}    
\end{equation}

Similar to traditional psychometric standards like Cronbach’s $\alpha$~\cite{cronbach1951coefficient}. Brennan’s~\cite{brennan2010generalizability} framework associates G-coefficients above 0.80 with strong reliability, supporting their use in tracking student engagement over time. In contrast, values between 0.60 and 0.80 indicate moderate reliability, useful for group-level trends but requiring caution for individual inferences. Similarly, the $\phi$-coefficient evaluates decision consistency, with values above 0.70 commonly used for distinguishing engagement patterns. A high $\phi$-coefficients for time-based measures indicates stable behavioral trends suitable for long-term monitoring.

\subsubsection{Predictive Validity of Measures (RQ2):}
We assess the predictive performance of each measure individually by estimating teacher-assigned year-end math scores while controlling for prior-year math scores. 

We follow Raftery's~\cite{raftery1995bayesian} guidelines for Bayesian Information Criterion (BIC) to evaluate the model performance: a BIC difference of $0\text{-}2$ indicates `weak'; $2\text{-}6$ `positive'; $6\text{-}10$ `strong'; and $>10$ `very strong' evidence in favor of the model with the smaller BIC. 

In the following equations, $j$ represents the student and $m$ represents the measure.

The baseline model establishes the relationship between the final math score ($Y_j$) and the prior math score ($X_j^{\text{prior}}$), as defined in Equation~\ref{eqn:baseline}. 

\begin{equation}
\label{eqn:baseline}
    Y_j = \beta_0 + \beta_1 X_j^{\text{prior}} + \epsilon
\end{equation}

We examine the predictive validity of each time-based measure ($X_j^{\text{tb}_m}$) listed in Table~\ref{tab:indicators_learner_behavior} as well as the students' gaming tendency by adding them to the baseline equation individually, Equation~\ref{eqn:individual_estimation_model}.

\begin{equation}
\label{eqn:individual_estimation_model}
    Y_j = \beta_0 + \beta_1 X_j^{\text{prior}} + \beta_2 X_j^{\text{tb}_{m}} + \epsilon
\end{equation}

Finally, we use Equation~\ref{eqn:combined_estimation_model} to identify the optimal set of measures ($X_{i}^{\text{tb}_1}, X_{i}^{\text{tb}_2}, \dots, X_{i}^{\text{tb}_m}$) to predict the students' final math score using step-wise regression with BIC for both forward and backward selection. This approach iteratively adds or removes predictors to optimize model fit. Additionally, we examine the Variance Inflation Factor (VIF)~\cite{fox2018r} of the final model to check for potential collinearity.

\begin{equation}
\label{eqn:combined_estimation_model}
    Y_j = \beta_0 + \beta_1 X_j^{\text{prior}} + \sum \beta_m X_{j}^{\text{tb}_{m}} + \epsilon
\end{equation}

This approach allows for the identification of the most informative set of measures beyond prior performance.

\subsubsection{Cross-System Generalizability of Measures \newline (RQ3):}
Finally, we use dataset 2 to assess the cross-system generalizability of the session-level time-based measures across DLPs. Similar to dataset 1, we computed the session-level measures and implemented the final model from the stepwise regression analysis, \textbf{RQ2} (Equation~\ref{eqn:combined_estimation_model}), to the student data from dataset 2. Details on dataset 2 were provided in Section~\ref{sec:Dataset}. A strong predictive performance here would attest to the robustness and generalizability of these time-based measures.

Stepwise regression can produce biased or unstable models, particularly when paired with lenient selection criteria or weakly motivated predictors. As such, we implement two safeguards in our analysis (\textbf{RQ2}). First, we use BIC\textemdash a relatively conservative model selection criterion known for favoring parsimonious models and recovering the true model structure under appropriate conditions. Second, we examine the final model using VIF to assess multicollinearity, ensuring that retained predictors contribute distinct and non-redundant information. These choices help mitigate common pitfalls associated with stepwise procedures. To further strengthen the validity of the selected features, we assess their cross-system generalizability in a different learning context (\textbf{RQ3}).

\section{Results}
As different classes met at different frequencies, we aggregated the data on a monthly basis to ensure consistent comparisons across students over time. The descriptive statistics for the monthly time-based measures aggregated across 9 months of the school year are presented in Table~\ref{tab:threshold_descriptives}. 

\begin{table}[!h]
\centering
\caption{Descriptive statistics of the monthly averages aggregated across 9 months. Direct time-based measures are reported in minutes, and the relative measures are z-scores.}
\label{tab:threshold_descriptives}
\resizebox{0.89\columnwidth}{!}{%
\begin{tabular}{lccc}
\hline
\multicolumn{1}{c}{Measure}&  \multicolumn{1}{c}{Mean}      & \multicolumn{1}{c}{SD} & \multicolumn{1}{c}{Median}\\ \hline
delayed start ($mins$)                  &     7.71      &  2.87    &   7.31    \\
relative delayed start                  &     0.03      &  0.42    &  -0.04     \\
session length ($mins$)                 &     29.34     &  4.36    &  30.01     \\
relative session length                 &    -0.04      &  0.41    &  -0.00     \\
early stop ($mins$)                     &     9.23      &  1.90    &   8.97     \\
relative early stop                     &     0.01      &  0.28    &  -0.00     \\
idle time ($> 2 mins$)                  &     4.18      &  2.08    &   3.84     \\
relative idle time                      &     0.01      &  0.36    &  -0.03     \\
total time on task ($mins$)             &     151.81    & 72.42    &  161.64    \\
usage time ratio                        &     0.66      &  0.08    &  0.66       \\
relative usage time ratio               &    -0.01      &  0.56    &   0.09      \\
attendance                              &     4.79      &  1.78    &  5.11       \\
gaming tendency                         &     0.09      &  0.37    &   0.075     \\
\hline
\end{tabular}}
\end{table}

\subsection{Stability of Measures (RQ1)}
\label{sec:stability_reliability_analysis}
The results of our first research question \textbf{RQ1} exploring the reliability of time-based measures are presented in Table~\ref{tab:reliability_metrics}. 

Session-level measures such as \textit{ delayed start} ($G = 0.76$), \textit{session length} ($G = 0.85$), and \textit{idle time} ($G = 0.81$) exhibited moderate to high reliability, indicating that these measures capture stable individual differences. A portion of the unexplained variance may still be systematic and potentially influenced by contextual factors such as task difficulty, classroom environment, holidays, or seasonal change. Provided these measures exhibit predictive validity (later in \textbf{RQ2}), the $\phi$-coefficients ($\phi = 0.75$–$0.84$) suggest that these measures can reliably distinguish individuals in observational studies or classroom interventions. 

\begin{table}[!ht]
    \centering
    \caption{Beside \textit{early-stop}, most session-level measures, exhibited moderate ($60 < G \leq 80$) to strong reliability ($G > 80 $). Notably, the reliability of most session-level measures was as good or better than gaming tendency.}
    \label{tab:reliability_metrics}
    \resizebox{0.7\columnwidth}{!}{%
    \begin{tabular}{lcc}
    \hline
    \multicolumn{1}{c}{Measure}                   & G          & $\phi$ \\ \hline
    delayed start ($mins$)                        & 0.76       & 0.75   \\
    relative delayed start                        & 0.82       & 0.82   \\
    session length ($mins$)                       & 0.85       & 0.84   \\
    relative session length                       & 0.81       & 0.81   \\
    early stop ($mins$)                           & 0.51       & 0.47   \\
    relative early stop                           & 0.55       & 0.55   \\
    idle time ($> 2 mins$)                        & 0.81       & 0.77   \\
    avg relative idle time                        & 0.73       & 0.73   \\
    total time on task ($mins$)                   & 0.94       & 0.90   \\
    usage time ratio                              & 0.76       & 0.73   \\
    relative time usage ratio                     & 0.76       & 0.76   \\
    attendance                                    & 0.93       & 0.90   \\
    gaming tendency                               & 0.76       & 0.76   \\ \hline
    \end{tabular}}
\end{table}

Unlike other session-level measures, \textit{early stop} exhibited much lower reliability ($G = 0.51$, $\phi = 0.47$), indicating significant variability. This suggests that \textit{early stop} is sensitive to transient or situational factors, such as class schedules, task completion, or teacher instructions. As such, it may have limited utility for student-level decision-making by itself.

Compared to session-level measures, more established measures such as \textit{total time on task} ($G = 0.94$) and \textit{attendance} ($G = 0.93$) demonstrated very strong reliability, whereas gaming tendency ($G = 0.76$) had moderate reliability. The very strong reliability of \textit{total time on task} and \textit{attendance} indicate that these behaviors remain consistent over time.  

Our exploration of the reliability of time-based measures \textbf{(RQ1)} reveals that session-level measures \textit{delayed start} and \textit{session length} and their relative estimates exhibited moderate to high reliability. In contrast, \textit{early stop} and its relative estimate were less reliable. These findings suggest that the session-level measures ($G > 0.80$) can be attributed to consistent individual differences across students, making them suitable for longitudinal analyses.  

\subsection{Predictive Validity of Measures (RQ2)}
\label{sec:predictive_validity}
 Given the reliability of most session-level measures, for \textbf{RQ2}, we aggregate them across the 9 months at the student level to examine their predictive validity. We standardize the final math score (mean = 84.20, SD = 8.03), prior math scores (mean = 84.27,  SD = 10.33), and the measures for interpretability. As outlined in Subsection~\ref{subsection:analysis-plan}, we use a combination of BIC and $R^2$ for the evaluation.

\subsubsection{Individual Predictive Performance}
\label{sec:predictive_validity}
The results comparing the individual predictive utility of the measures are presented in Table~\ref{tab:predictive-validity-model}. Incorporating certain session-level measures into the baseline model resulted in strong performance improvements, i.e., a reduction in BIC ($>10$). Specifically, we observed strong evidence of performance improvements for \textit{delayed start} (BIC = 380, $R^2$ = 0.51), \textit{relative delayed start} (BIC = 376, $R^2$ = 0.53), \textit{relative session length} (BIC = 383, $R^2$ = 0.50), \textit{idle time} (BIC = 386, $R^2$ = 0.49), \textit{relative idle time} (BIC = 390, $R^2$ = 0.48), and \textit{relative time usage ratio} (BIC = 382, $R^2$ = 0.50).

\begin{table}[!ht]
    \centering
    \caption{Examining the predictive validity of the session-level measures. The measures that demonstrated strong evidence (i.e., reduction of the BIC $> 10$~\cite{raftery1995bayesian}) are bolded in the table.}
    \label{tab:predictive-validity-model}
    \resizebox{\columnwidth}{!}{%
    \begin{tabular}{l S[table-format=-1.2,table-space-text-post={***}] c c}
    \hline
    \multicolumn{1}{c}{Model}               & $\beta$                             & $R^2$   & BIC          \\ \hline
    baseline                                & \textemdash                         & 0.44    & 401          \\
    baseline + delayed start ($mins$)       & -0.23\textsuperscript{\textbf{***}} & 0.51    & \textbf{380} \\
    baseline + relative delayed start       & -0.25\textsuperscript{\textbf{***}} & 0.52    & \textbf{376} \\
    baseline + session length ($mins$)      & 0.13\textsuperscript{\textbf{**}}   & 0.46    & 398          \\
    baseline + relative session length      & 0.23\textsuperscript{\textbf{***}}  & 0.50    & \textbf{383} \\
    baseline + early stop ($mins$)          & 0.04                                & 0.44    & 406          \\
    baseline + relative early stop          & 0.07                                & 0.44    & 404          \\
    baseline + idle time ($> 2 mins$)       & -0.21\textsuperscript{\textbf{***}} & 0.49    & \textbf{386} \\
    baseline + relative idle time           & -0.18\textsuperscript{\textbf{***}} & 0.48    & \textbf{390} \\
    baseline + total time on task ($mins$)  & 0.02                                & 0.44    & 406          \\
    baseline + usage time ratio             & 0.11\textsuperscript{\textbf{*}}    & 0.45    & 400          \\
    baseline + relative time usage ratio    & 0.22\textsuperscript{\textbf{***}}  & 0.50    & \textbf{382} \\
    baseline + attendance                   & -0.02                               & 0.44    & 406          \\
    baseline + gaming tendency              & -0.12\textsuperscript{\textbf{**}}  & 0.46    & 399          \\ \hline
    \multicolumn{4}{l}{\scriptsize $\textsuperscript{*}p < 0.05$, $\textsuperscript{**}p < 0.01$, $\textsuperscript{***}p < 0.001$}
    \end{tabular}}
\end{table}

\textit{Session length, usage time ratio}, and \textit{gaming tendency} were also significant predictors of final math scores. However, the performance improvements to the baseline model were relatively weak. Finally, despite their very strong reliability, \textit{total time on task} and \textit{attendance} did not improve the performance of the baseline model.

\subsubsection{Combined Predictive Performance}
Next, we use stepwise regression to examine the combined performance of the individual measures to explore how the new session-level measures combine with other measures to enhance the prediction of student learning outcomes.

We observed a strong Pearson's correlation between: \textit{total time on task} and \textit{attendance} ($r = 0.96, p < 0.001$), as well as \textit{relative delayed start} and \textit{delayed start} ($r = 0.95, p < 0.001$). As such, we only included \textit{total time on task} and \textit{relative delayed start} (measures with lower BIC in Table~\ref{tab:predictive-validity-model}) in the final analysis to avoid potential multicollinearity.

The results for both forward and backward stepwise regression are presented in Table~\ref{tab:combined-predictive-validity-model}. The forward stepwise regression model achieved an $R^2$ of 0.53 and a BIC of 373. From the set of available measures forward stepwise regression included~\textit{relative delayed start} ($\beta = -0.25, p < 0.001$), and \textit{gaming tendency} ($\beta = -0.12, p = 0.005$) in the final model and both measures had a significant negative correlation with final math score. 

Similarly, the backward stepwise regression model achieved an $R^2$ of 0.55 and a BIC of 372. From the set of available measures backward stepwise regression included~\textit{Relative delayed start} ($\beta = -0.21, p < 0.001$), \textit{early stop} ($\beta = -0.13, p = 0.003$), and \textit{idle time} ($\beta = -0.16, p = 0.002$) in the final model and all three measures had a significant negative correlation with final math score. 

\begin{table}[h]
    \centering
    \caption{Combined predictive performance of time-based measures using BIC-based stepwise regression.}
    \label{tab:combined-predictive-validity-model}
    \resizebox{0.95\columnwidth}{!}{%
    \begin{tabular}{l S[table-format=-1.2,table-space-text-post={***}] S[table-format=-1.2,table-space-text-post={***}]}
    \hline
                                    &  \multicolumn{2}{c}{final math score}  \\ \cline{2-3}
    \multicolumn{1}{c}{Predictor}   & {$\beta$ (forward)}                  & {$\beta$ (backward)}                   \\ \hline
    Intercept                       & 0.03                                 & 0.03                                 \\
    relative delayed start          & -0.25\textsuperscript{\textbf{***}}  & -0.21\textsuperscript{\textbf{***}}  \\
    early stop ($mins$)             &   \textemdash                        & -0.13\textsuperscript{\textbf{**}}   \\
    idle time ($> 2 mins$)          &   \textemdash                        & -0.16\textsuperscript{\textbf{**}}  \\
    prior final score               & 0.47\textsuperscript{\textbf{***}}   & 0.45\textsuperscript{\textbf{***}}   \\ 
    gaming tendency                 & -0.12\textsuperscript{\textbf{**}}   &  \textemdash                         \\ \hline
    \multicolumn{2}{l}{\scriptsize $\textsuperscript{*}p < 0.05$, $\textsuperscript{**}p < 0.01$, $\textsuperscript{***}p < 0.001$}
    \end{tabular}}
\end{table}

For \textbf{RQ2}, our exploration of the predictive performance of individual measures reveals that session-level measures \textit{relative delayed start} was the most predictive of student final math score across all explored measures. Similarly, when exploring the combined predictive validity of these measures, \textit{relative delayed start} was selected in the best-performing models for both forward and backward stepwise regression.  

\subsection{Cross-System Generalizability of Measures (RQ3)}
For \textbf{RQ3}, we assess the generalizability of session-level time-based measures in a new context and learning environment. We use session-level measures from i-Ready data to predict student performance on a standardized state test. 

Compared to Cognitive Tutor, i-Ready had a higher average transaction duration of 1.87 minutes (SD = 41.98 seconds). As such, we set the threshold for idle time to $>$ 3.94 minutes ($mean + 3 * SD$). Additionally, we exclude \textit{gaming tendency} due to the lack of an equivalent measure in i-Ready.

The time-based measures of students using i-Ready exhibited skewed distributions: \textit{early stop} ($M = 4.20$, $SD = 4.62$) and \textit{idle time} ($M = 5.40$, $SD = 4.12$). Using the Fisher-Pearson standardized moment coefficient, \textit{idle time} ($g_1 = 1.40$) showed moderate right skew, while \textit{early stop} was highly skewed ($g_1 = 3.05$). To correct for skewness, we applied a square root transformation to \textit{idle time} and a log transformation to \textit{early stop}, reducing their skewness from 1.40 to 0.30 and 3.05 to 0.30, respectively.

The results examining the generalizability of the stepwise regression models from RQ2 for students using i-Ready are reported in Table~\ref{tab:generalizability-predictive-validity-model}. The replicated forward regression model achieved a BIC of 1526 and an $R^2$ of 0.52. \textit{Relative delayed start} ($\beta = -0.13, p < 0.001$) had a significant negative correlation with student performance on the state test.

Similarly, the replicated backward regression model had a BIC of 1530 and an $R^2$ of 0.53. \textit{Relative delayed start} ($\beta = -0.12, p < 0.001$) remained significantly negatively correlated with student performance. However, neither \textit{early stop} ($\beta = 0.04, p = 0.19$) nor \textit{idle time} ($\beta = -0.03, p = 0.30$) showed a significant relationship with student performance on the state test.
\newline 

\begin{table}[h]
    \centering
    \caption{Combined predictive performance of time-based measures using BIC-based stepwise regression in i-Ready data.}
    \label{tab:generalizability-predictive-validity-model}
    \resizebox{\columnwidth}{!}{%
    \begin{tabular}{l S[table-format=-1.2,table-space-text-post={***}] S[table-format=-1.2,table-space-text-post={***}]}
    \hline
                                          &  \multicolumn{2}{c}{state test score} \\ \cline{2-3}
    \multicolumn{1}{c}{Predictor}         & {$\beta$ (forward)}                  & {$\beta$ (backward)}                    \\ \hline
    Intercept                             & 0.00                                  & 0.00                                   \\
    relative delayed start                & -0.13\textsuperscript{\textbf{***}}   & -0.12\textsuperscript{\textbf{***}}    \\
    early stop ($transformed$)            &    \textemdash                        & 0.04                                   \\
    idle time ($transformed$)             &    \textemdash                        & -0.03                                  \\
    prior state test score                & 0.67\textsuperscript{\textbf{***}}    & 0.67\textsuperscript{\textbf{***}}     \\ \hline
    \multicolumn{2}{l}{\scriptsize $\textsuperscript{*}p < 0.05$, $\textsuperscript{**}p < 0.01$, $\textsuperscript{***}p < 0.001$}
    \end{tabular}}
\end{table}

For \textbf{RQ3}, our exploration of the cross-system generalizability of session-level measures demonstrates that \textit{relative delayed start} remains a robust predictor of student performance on standardized state tests. 
\newline 

\section{Discussion}
A core objective of EDM and Learning Analytics is to develop system-general measures that reliably capture learning behaviors across diverse learning environments. This study takes a step towards this broader goal by evaluating the predictive validity and generalizability of naturally embedded session-level engagement measures that provide insights into how student regulate their behavior.
\subsection{Findings}

\subsubsection{Surprisingly Strong Predictive Performance of\newline Session-Level Measures}
We evaluate the temporal stability (\textbf{RQ1}) of session-level measures using G-Theory~\cite{cronbach1972dependability,brennan2010generalizability}. The results indicate that session-level time-based measures capture stable aspects of student behavior—many exhibiting reliability comparable to or exceeding that of \textit{gaming tendency} ($G=0.76,\phi=0.76$). While session-level measures showed lower reliability than global measures such as total \textit{time on task} ($G=0.94,\phi=0.90$) and \textit{attendance} ($G=0.93,\phi=0.90$), all measures, except for \textit{early stop} ($G=0.51,\phi=0.47$) and \textit{relative early stop} ($G=0.55,\phi=0.55$), exhibited moderate to strong reliability  ($G\geq0.70,\phi=0.70$). 

Next, given their reliability, we evaluate the predictive validity (\textbf{RQ2}) of individual session-level measures in predicting student learning outcomes. Specifically, \textit{delayed start, relative delayed start, and relative session length} outperformed more well-established measures such as \textit{time on task, attendance}, and \textit{gaming} by demonstrating stronger improvements to the performance of the baseline model (Table~\ref{tab:predictive-validity-model}). 

Taken together, the stability (\textbf{RQ1}) and predictive validity (\textbf{RQ2}) of individual session-level measures (Table~\ref{tab:predictive-validity-model}) provide valuable insight into student engagement. For example, on average, the classwork session lasted for 45 minutes (Table~\ref{tab:descriptive_statistics_classwork}) with students \textit{delayed starting} by approximately 8 minutes and spending about 29 minutes actively working on their math (Table~\ref{tab:threshold_descriptives}: session length). So even when the student \textit{delayed starting} their work, they had enough time to make up for their \textit{delayed start}. Yet \textit{delayed start} was still a stronger predictor of learning outcomes than \textit{time on task} and individual \textit{session length}. This indicates that \textit{delayed start} captures aspects of learner behavior that go beyond time, i.e., students who start early (low \textit{delayed start}) tend to be more engaged, manage time more effectively, and ultimately achieve better learning outcomes.

The combined predictive validity analysis (\textbf{RQ2}: Table~\ref{tab:combined-predictive-validity-model}) illustrates how effectively session-level measures can be incorporated with process-level measures (e.g., \textit{gaming tendency}) to enhance the prediction of student learning outcomes. When predicting learning outcomes (Table~\ref{tab:combined-predictive-validity-model}), we find standardized \textit{relative delayed start} ($\beta = -0.25$) to be twice as predictive as standardized \textit{gaming tendency} ($\beta = -0.12$) and approximately half as predictive as standardized \textit{prior final grade} ($\beta = 0.47$). Our findings regarding the cross-system generalizability (\textbf{RQ3}) of these session-level measures further reinforce the robustness and utility value of \textit{relative delayed start} and \textit{early stop} in predicting student learning outcomes. This is particularly notable given that the rule-based gaming detector requires some effort to be implemented in new systems and is currently limited to two systems, MATHia~\cite{ritter2007cognitive} (formerly Cognitive Tutor) and ASSISTments~\cite{heffernan2014assistments}.

\subsubsection{Contextual Sensitivity: A More Meaningful\newline Set of Indicators Than Time on Task}
Beyond their predictive performance, results from \textbf{RQ1} and \textbf{RQ2} reveal a trade-off between reliability and sensitivity to change for certain session-level measures. Although their reliability ($G>0.75$) is lower than \textit{time on task} ($G=94$), session-level measures (i.e., \textit{delayed start, session length}) significantly outperformed \textit{time on task} in predicting learning outcomes. This suggests that session-level measures are more sensitive to contextual factors that influence learning. As such, \textit{delayed start} may serve as a stronger standalone outcome or a complementary measure for identifying differential intervention effects in studies using increased \textit{time on task}~\cite{Nagashima2022} as an outcome.

\subsubsection{Reliability Comparable to Psychometric\newline Measures}
To contextualize the reliability and predictive validity of session-level measures more broadly, psychometric measures such as the Big Five personality traits typically achieve \textit{G}-coefficients between 0.80 and 0.90 and \textit{$\phi$}-coefficients between 0.75 and 0.85~\cite{arterberry2014application}. The Big Five personality traits are extensively validated against academic outcomes~\cite{paunonen2001big} and are generally considered stable over time~\cite{cobb2012stability}, making them a suitable reference for evaluating the reliability of our session-level measures. Among The Big Five, conscientiousness demonstrates the strongest predictive validity for academic achievement~\cite{dumfart2016conscientiousness}. When comparing the G and $\phi$ coefficients of \textit{relative delayed start} with \textit{conscientiousness}, from Arterberry et al.~\cite{arterberry2014application}, we find the G-coefficient of \textit{relative delayed start} at 0.82 closely matches \textit{conscientiousness} at 0.84 and the $\phi$-coefficient at 0.82 exceeds \textit{conscientiousness} at 0.73. This finding is both surprising and encouraging and highlights how consistent students are in terms of delaying to start work with respect to their peers, with those starting early benefiting the most from this behavior in the long run (based on our findings form \textbf{RQ2} and \textbf{RQ3}). Our findings align with prior works reporting procrastination differences between students to be generally stable and highly correlated with other personality traits related to academic achievement (i.e., conscientiousness)~\cite{lay1998relation}. 

Finally, it is worth reiterating that the predictive validity of session-level measures was demonstrated across two different DLPs and two distinct outcome measures. Teacher-assigned final grades with Cognitive Tutor data (\textbf{RQ1} and \textbf{RQ2}) reflect a more comprehensive evaluation of student effort, participation, and coursework. In contrast, standardized state test scores with i-Ready data (\textbf{RQ3}) provide a more uniform assessment of student proficiency. The fact that session-level measures predict both underscores their broad relevance.

\subsection{Broader Implications in EDM}
\subsubsection{Classwork Sessions and Student Engagement}
Developing reliable strategies to infer learning sessions has been a persistent challenge in EDM, with session boundaries often characterized as a ``black box'' due to the difficulty of distinguishing when sessions begin and end~\cite{kovanovic2015penetrating}. However, our analysis across two learning platforms suggests that classwork sessions in K-12 settings tend to follow a structured and predictable engagement pattern (Figure~\ref{fig:mathia-threshold-time-session}).

As demonstrated in Section~\ref{subsubsec:student_class_session}, classwork sessions in K-12 settings follow clearer expectations for learner engagement than those in higher education, where students work independently or use learning platforms for homework—resulting in more fluid session boundaries~\cite{gasevic2017detecting,fan2021learning,liu2015modeling,kizilcec2017self}. This structured engagement in K-12 classrooms allows for a straightforward yet effective way to define session boundaries, making it possible to develop a scalable session inference method across different learning environments. Our approach (Algorithm~\ref{algo:class_inference_algo}) is deliberately simple, requiring minimal calibration while remaining highly adaptable across platforms. Specifically, the algorithm can be applied to any system that logs student engagement data—a feature common to virtually all learning platforms, including MATHia~\cite{ritter2016mathia}, ASSISTments~\cite{heffernan2014assistments}, Graspable Maths~\cite{weitnauer2016graspable}, and IXL~\cite{an2024impact}. By providing an easy-to-implement yet powerful session inference method, this approach aims to open new avenues for investigating behavioral dynamics and their long-term influence on learning.

\subsubsection{Toward Scalable and Generalizable Measures}
As highlighted in Baker’s Challenges for the Future of Learning Analytics and EDM~\cite{baker2019challenges}, developing detectors that capture learning behaviors that are both generalizable and predictive of learning outcomes is a key focus in EDM. Prior research has applied knowledge- and feature-engineering methods to detect learner behaviors such as gaming~\cite{paquette2014towards}, and effort~\cite{gurung2021examining,shih2011response}. Our work continues this line of research by leveraging feature-engineered methods to develop measures that offer more nuanced insights into student engagement and learning. This simple approach (Algorithm~\ref{algo:class_inference_algo}) for dynamically inferring classwork sessions creates new opportunities in EDM for studying session-level differences within and across students. These session-level measures present new opportunities to refine predictive models for affect~\cite{botelho2018studying,botelho2019machine,d2012dynamics}, behavior~\cite{baker2004off,christoff2018mind,mills2016automatic,beck2013wheel} and learning~\cite{pardos2013affective}. Furthermore, these measures provide a more context-aware representation of student engagement and behavior while being intuitive, interpretable, and aligned with classroom norms and teacher expectations, a central focus of EDM.

\section{Limitations}
Although the current Algorithm~\ref{algo:class_inference_algo} is promising, it remains sensitive to outliers. A single student who continues working beyond typical class times can skew session-level measures, especially when most of the class has stopped. Incorporating additional information—such as the density of active students relative to session attendees—could improve the algorithm’s precision. This improvement in session end-time inference would also help us potentially identify new groups of students who work beyond the bell schedule or their peers. Similar to early starters (low \textit{delayed start}), persistent engagement beyond their peers could also reveal valuable insights into learner motivation and goal orientation~\cite{gollwitzer2006implementation}. For instance, are these students desperate to catch up? Do they enjoy doing math? Or are they working ahead?

Session-level time-based measures offer valuable insights into student engagement and provide opportunities for future research. While behaviors such as \textit{delayed start, session length, and early stopping} likely reflect self-regulatory (e.g., goal-setting, strategic planning) and motivational (e.g., diligence, persistence) factors, further investigation is needed to understand how these constructs manifest in session-level data. For instance, a high \textit{delayed start} may indicate procrastination~\cite{borchers2024you}, low self-efficacy~\cite{winne2017cognition,wolters2003regulation}, or poor time management~\cite{park2018understanding,wolters2021college}. Likewise, performance-avoidance tendencies may delay student engagement to avoid failure~\cite{midgley2001performance,gollwitzer2006implementation}, whereas students with mastery goals may start earlier to optimize learning opportunities~\cite{dweck1986motivational}.

\section{Future Research}
To further strengthen the validity of session-level measures, future research should examine their construct validity by exploring their relationship with survey-based measures of motivational and self-regulatory constructs such as SRL~\cite{pintrich2004conceptual}, diligence~\cite{galla2014academic}, and conscientiousness~\cite{kim2016conscientiousness}. Dang et al.\cite{dang2017detecting} found that \textit{time on task} influences diligence; next, research should explore how session-level behaviors shape \textit{time on task} and, in turn, productive engagement. Incorporating classroom observations\cite{ocumpaugh2015baker}, teacher insights~\cite{holstein2019co}, and frequent student polling using survey measures as prompts~\cite{bernacki2013datasource} can further contextualize these behaviors. Integrating session-level measures into EDM and Learning Analytics frameworks can enhance understanding of underlying student behaviors such as motivation~\cite{wolters2021college} and goal setting~\cite{dweck1986motivational,midgley2001performance}. Additionally, these insights can inform the development of teaching augmentation tools and dashboards~\cite{holstein2018informing,an2020ta}, where the real benefit lies in how closely these measures align with teacher expectations—such as the intuitive heuristics educators use to distinguish diligent students, particularly those who start early versus those who delay or sustain engagement for longer (\textit{session length}) versus those who stop early.


Another promising direction is incorporating session-level information into affect and behavioral modeling, particularly in settings where learning platform usage is heavily classwork-based. As generalizability remains a key challenge in EDM~\cite{baker2019challenges}, session-level measures offer a new perspective on engagement dynamics within structured classroom environments. For example, session-level measures could help contextualize aspects of student affect dynamics~\cite{d2012dynamics, botelho2018studying}. Students who consistently delay starting work and stop early may experience heightened frustration while working on problems, or their behavior may reflect a lack of interest in the subject, perceiving it as boring. By situating learning behaviors within the structured constraints of classroom sessions, these measures can provide deeper insights into session-level measures and help explain student affinity to transition from one affective state to the next. We present affective modeling as one potential application. However, session-level measures have a broader potential to contribute to user modeling in various learning environments.

Finally, improving the precision of session boundary detection remains an important area for future work. While our current approach (Algorithm~\ref{algo:class_inference_algo}) infers sessions from log data, further validation is needed to assess its performance. Incorporating known bell schedules, when available, could anchor session boundaries to classroom norms and provide a more structured baseline for comparison. For instance, if a subset of highly motivated students routinely start early or stay beyond class hours, students who begin and end work on time may be incorrectly labeled as delayed starters or early stoppers. Beyond these misclassifications, the motivational significance of the highly motivated students also goes undetected. Comparing the algorithm’s performance against actual classroom schedules would help refine the algorithm's performance. This, in turn, could help distinguish meaningful behavioral patterns from systemic factors, ultimately enhancing the insights provided by session-level measures. 

\section{Conclusion}
Our examination of session-level measures of student practice in math software shows that these measures are surprisingly reliable and valid for predicting learning outcomes. Notably, \textit{relative delayed start} emerged as the strongest session-level predictor\textemdash twice as predictive as \textit{gaming-the-system behavior} and half as predictive as prior performance. Furthermore, \textit{delayed start, relative delayed start} and \textit{relative session length} were predictive beyond \textit{total time on task}. Session-level measures also demonstrated cross-system generalizability, transferring effectively between platforms (Cognitive Tutor to i-Ready) and across different learning outcomes (teacher-assigned final grades to standardized state tests). We attribute the cross-system generalizability and predictive validity of these measures to their temporal stability, suggesting they capture stable and meaningful differences in student behavior. These session-level measures also offer a naturally embedded complement to survey-based engagement metrics, which rely on self-reporting and may not always reflect actual learning behaviors. Unlike surveys, they capture what students do rather than what they say, leveraging existing log data without additional instrumentation or disruption to instruction. By providing scalable, behaviorally grounded indicators of engagement, session-level measures offer a valuable tool for understanding self-regulated student learning.

In EDM, gaming the system is among the few detectors with a well-established track record of cross-system generalizability and predictive validity. This study adds to the set of generalizable detectors by introducing a new class of session-level measures, which are system-agnostic and broadly applicable across diverse learning platforms. 
More broadly, these findings demonstrate the potential of EDM methods in extending preexisting session-level measures and establishing reliable indicators of student academic performance. Furthermore, these measures have the potential to further our understanding of learner behavior by providing valuable insight into aspects of student motivation, effort, and self-regulation. 

\section*{Acknowledgments}
The authors would like to thank Curriculum Associates, LLC for their partnership and collaborative efforts in this research. This work was supported by the Learning Engineering Virtual Institute. The opinions, findings, and conclusions expressed in this material are those of the authors and do not necessarily reflect the views of the Institute or Curriculum Associates.

We also wish to acknowledge Matthew L. Bernacki for his efforts in collecting the original Cognitive Tutor dataset used in this study, and Steven Dang, whose dissertation work on session-level measures and diligence were foundational to this work.
%
\bibliographystyle{abbrv}
\bibliography{EDM_Article_Submission}  

\begin{thebibliography}{10}

\bibitem{an2020ta}
P.~An, K.~Holstein, B.~d'Anjou, B.~Eggen, and S.~Bakker.
\newblock The ta framework: Designing real-time teaching augmentation for k-12 classrooms.
\newblock In {\em Proceedings of the 2020 CHI Conference on Human Factors in Computing Systems}, pages 1--17, 2020.

\bibitem{an2024impact}
X.~An and C.~Schonberg.
\newblock The impact of ixl on students’ math self-efficacy, 2024.

\bibitem{arterberry2014application}
B.~J. Arterberry, M.~P. Martens, J.~M. Cadigan, and D.~Rohrer.
\newblock Application of generalizability theory to the big five inventory.
\newblock {\em Personality and individual differences}, 69:98--103, 2014.

\bibitem{azevedo2013metacognition}
R.~Azevedo and V.~Aleven.
\newblock Metacognition and learning technologies: An overview of current interdisciplinary research.
\newblock {\em International handbook of metacognition and learning technologies}, pages 1--16, 2013.

\bibitem{baker2004off}
R.~S. Baker, A.~T. Corbett, K.~R. Koedinger, and A.~Z. Wagner.
\newblock Off-task behavior in the cognitive tutor classroom: When students" game the system".
\newblock In {\em Proceedings of the SIGCHI conference on Human factors in computing systems}, pages 383--390, 2004.

\bibitem{baker2019challenges}
R.~S. Baker et~al.
\newblock Challenges for the future of educational data mining: The baker learning analytics prizes.
\newblock {\em Journal of educational data mining}, 11(1):1--17, 2019.

\bibitem{baker2008developing}
R.~S.~d. Baker, A.~T. Corbett, I.~Roll, and K.~R. Koedinger.
\newblock Developing a generalizable detector of when students game the system.
\newblock {\em User Modeling and User-Adapted Interaction}, 18:287--314, 2008.

\bibitem{beck2013wheel}
J.~E. Beck and Y.~Gong.
\newblock Wheel-spinning: Students who fail to master a skill.
\newblock In {\em Artificial Intelligence in Education: 16th International Conference, AIED 2013, Memphis, TN, USA, July 9-13, 2013. Proceedings 16}, pages 431--440. Springer, 2013.

\bibitem{bernacki2013datasource}
M.~Bernacki and S.~Ritter.
\newblock Hopewell 2011-2012 algebra (learning orientation survey).
\newblock Dataset 718 in DataShop, 2013.
\newblock Retrieved from.

\bibitem{bernard1993diligence}
H.~Bernard, J.~D. Thayer, and E.~A. Streeter.
\newblock Diligence and academic performance.
\newblock {\em Journal of Research on Christian Education}, 2(2):213--234, 1993.

\bibitem{borchers2024you}
C.~Borchers, Y.~Xu, and Z.~A. Pardos.
\newblock Are you an early dropper or late shopper? mining enrollment transaction data to study procrastination in higher education.
\newblock In {\em Proceedings of the 17th International Conference on Educational Data Mining}, pages 426--433, 2024.

\bibitem{borchers2025workload}
C.~Borchers, Y.~Xu, Z.~A. Pardos, et~al.
\newblock Workload overload? late enrollment leads to course dropout.
\newblock {\em Journal of Educational Data Mining}, 17(1):126--156, 2025.

\bibitem{botelho2019machine}
A.~F. Botelho, R.~S. Baker, and N.~T. Heffernan.
\newblock Machine-learned or expert-engineered features? exploring feature engineering methods in detectors of student behavior and affect.
\newblock In {\em The twelfth international conference on educational data mining}, 2019.

\bibitem{botelho2018studying}
A.~F. Botelho, R.~S. Baker, J.~Ocumpaugh, and N.~T. Heffernan.
\newblock Studying affect dynamics and chronometry using sensor-free detectors.
\newblock {\em International Educational Data Mining Society}, 2018.

\bibitem{brennan1992generalizability}
R.~L. Brennan.
\newblock Generalizability theory.
\newblock {\em Educational Measurement: Issues and Practice}, 11(4):27--34, 1992.

\bibitem{brennan2010generalizability}
R.~L. Brennan.
\newblock Generalizability theory and classical test theory.
\newblock {\em Applied measurement in education}, 24(1):1--21, 2010.

\bibitem{brookfield2015skillful}
S.~D. Brookfield.
\newblock {\em The skillful teacher: On technique, trust, and responsiveness in the classroom}.
\newblock John Wiley \& Sons, 2015.

\bibitem{christoff2018mind}
K.~Christoff, C.~Mills, J.~R. Andrews-Hanna, Z.~C. Irving, E.~Thompson, K.~C. Fox, and J.~W. Kam.
\newblock Mind-wandering as a scientific concept: cutting through the definitional haze.
\newblock {\em Trends in cognitive sciences}, 22(11):957--959, 2018.

\bibitem{cobb2012stability}
D.~A. Cobb-Clark and S.~Schurer.
\newblock The stability of big-five personality traits.
\newblock {\em Economics Letters}, 115(1):11--15, 2012.

\bibitem{cocea2009impact}
M.~Cocea, A.~Hershkovitz, and R.~S. Baker.
\newblock The impact of off-task and gaming behaviors on learning: immediate or aggregate?
\newblock In {\em Artificial Intelligence in Education}, pages 507--514. Ios Press, 2009.

\bibitem{cronbach1951coefficient}
L.~J. Cronbach.
\newblock Coefficient alpha and the internal structure of tests.
\newblock {\em psychometrika}, 16(3):297--334, 1951.

\bibitem{cronbach1972dependability}
L.~J. Cronbach, G.~C. Gleser, H.~Nanda, and N.~Rajaratnam.
\newblock {\em The Dependability of Behavioral Measurements: Theory of Generalizability for Scores and Profiles}.
\newblock Wiley, New York, NY, 1972.

\bibitem{dang2017detecting}
S.~Dang, M.~Yudelson, and K.~R. Koedinger.
\newblock Detecting diligence with online behaviors on intelligent tutoring systems.
\newblock In {\em Proceedings of the Fourth (2017) ACM Conference on Learning@ Scale}, pages 51--59, 2017.

\bibitem{dang2022disseration}
S.~C. Dang.
\newblock {\em Exploring Behavioral Measurement Models of Learner Motivation}.
\newblock PhD thesis, Carnegie Mellon University, 2022.

\bibitem{dang2020ebb}
S.~C. Dang and K.~R. Koedinger.
\newblock The ebb and flow of student engagement: Measuring motivation through temporal pattern of self-regulation.
\newblock {\em International Educational Data Mining Society}, 2020.

\bibitem{duckworth2007grit}
A.~L. Duckworth, C.~Peterson, M.~D. Matthews, and D.~R. Kelly.
\newblock Grit: perseverance and passion for long-term goals.
\newblock {\em Journal of personality and social psychology}, 92(6):1087, 2007.

\bibitem{dumfart2016conscientiousness}
B.~Dumfart and A.~C. Neubauer.
\newblock Conscientiousness is the most powerful noncognitive predictor of school achievement in adolescents.
\newblock {\em Journal of individual Differences}, 2016.

\bibitem{dweck1986motivational}
C.~S. Dweck.
\newblock Motivational processes affecting learning.
\newblock {\em American psychologist}, 41(10):1040, 1986.

\bibitem{d2012dynamics}
S.~D’Mello and A.~Graesser.
\newblock Dynamics of affective states during complex learning.
\newblock {\em Learning and Instruction}, 22(2):145--157, 2012.

\bibitem{eccles2020expectancy}
J.~S. Eccles and A.~Wigfield.
\newblock From expectancy-value theory to situated expectancy-value theory: A developmental, social cognitive, and sociocultural perspective on motivation.
\newblock {\em Contemporary educational psychology}, 61:101859, 2020.

\bibitem{fan2021learning}
Y.~Fan, W.~Matcha, N.~A. Uzir, Q.~Wang, and D.~Ga{\v{s}}evi{\'c}.
\newblock Learning analytics to reveal links between learning design and self-regulated learning.
\newblock {\em International Journal of Artificial Intelligence in Education}, 31(4):980--1021, 2021.

\bibitem{fox2018r}
J.~Fox and S.~Weisberg.
\newblock {\em An R companion to applied regression}.
\newblock Sage publications, 2018.

\bibitem{furrer2014influence}
C.~J. Furrer, E.~A. Skinner, and J.~R. Pitzer.
\newblock The influence of teacher and peer relationships on students’ classroom engagement and everyday motivational resilience.
\newblock {\em Teachers College Record}, 116(13):101--123, 2014.

\bibitem{galla2014academic}
B.~M. Galla, B.~D. Plummer, R.~E. White, D.~Meketon, S.~K. D'Mello, and A.~L. Duckworth.
\newblock The academic diligence task (adt): Assessing individual differences in effort on tedious but important schoolwork.
\newblock {\em Contemporary educational psychology}, 39(4):314--325, 2014.

\bibitem{gasevic2017detecting}
D.~Gasevic, J.~Jovanovic, A.~Pardo, and S.~Dawson.
\newblock Detecting learning strategies with analytics: Links with self-reported measures and academic performance.
\newblock {\em Journal of Learning Analytics}, 4(2):113--128, 2017.

\bibitem{gollwitzer2006implementation}
P.~M. Gollwitzer and P.~Sheeran.
\newblock Implementation intentions and goal achievement: A meta-analysis of effects and processes.
\newblock {\em Advances in experimental social psychology}, 38:69--119, 2006.

\bibitem{gong2015towards}
Y.~Gong and J.~E. Beck.
\newblock Towards detecting wheel-spinning: Future failure in mastery learning.
\newblock In {\em Proceedings of the second (2015) ACM conference on learning@ scale}, pages 67--74, 2015.

\bibitem{gurung2021examining}
A.~Gurung, A.~F. Botelho, and N.~T. Heffernan.
\newblock Examining student effort on help through response time decomposition.
\newblock In {\em LAK21: 11th International Learning Analytics and Knowledge Conference}, pages 292--301, 2021.

\bibitem{heffernan2014assistments}
N.~T. Heffernan and C.~L. Heffernan.
\newblock The assistments ecosystem: Building a platform that brings scientists and teachers together for minimally invasive research on human learning and teaching.
\newblock {\em International Journal of Artificial Intelligence in Education}, 24:470--497, 2014.

\bibitem{holstein2018informing}
K.~Holstein, B.~M. McLaren, and V.~Aleven.
\newblock Informing the design of teacher awareness tools through causal alignment analysis.
\newblock International Society of the Learning Sciences, Inc.[ISLS]., 2018.

\bibitem{holstein2018student}
K.~Holstein, B.~M. McLaren, and V.~Aleven.
\newblock Student learning benefits of a mixed-reality teacher awareness tool in ai-enhanced classrooms.
\newblock In {\em Artificial Intelligence in Education: 19th International Conference, AIED 2018, London, UK, June 27--30, 2018, Proceedings, Part I 19}, pages 154--168. Springer, 2018.

\bibitem{holstein2019co}
K.~Holstein, B.~M. McLaren, and V.~Aleven.
\newblock Co-designing a real-time classroom orchestration tool to support teacher-ai complementarity.
\newblock {\em Grantee Submission}, 2019.

\bibitem{huang2023using}
Y.~Huang, S.~Dang, J.~Elizabeth~Richey, P.~Chhabra, D.~R. Thomas, M.~W. Asher, N.~G. Lobczowski, E.~A. McLaughlin, J.~M. Harackiewicz, V.~Aleven, et~al.
\newblock Using latent variable models to make gaming-the-system detection robust to context variations.
\newblock {\em User Modeling and User-Adapted Interaction}, pages 1--47, 2023.

\bibitem{kim2016conscientiousness}
L.~E. Kim, A.~E. Poropat, and C.~MacCann.
\newblock Conscientiousness in education: Its conceptualization, assessment, and utility.
\newblock {\em Psychosocial skills and school systems in the 21st century: Theory, research, and practice}, pages 155--185, 2016.

\bibitem{kizilcec2017self}
R.~F. Kizilcec, M.~P{\'e}rez-Sanagust{\'\i}n, and J.~J. Maldonado.
\newblock Self-regulated learning strategies predict learner behavior and goal attainment in massive open online courses.
\newblock {\em Computers \& education}, 104:18--33, 2017.

\bibitem{kovanovic2015penetrating}
V.~Kovanovi{\'c}, D.~Ga{\v{s}}evi{\'c}, S.~Dawson, S.~Joksimovi{\'c}, R.~S. Baker, and M.~Hatala.
\newblock Penetrating the black box of time-on-task estimation.
\newblock In {\em Proceedings of the fifth international conference on learning analytics and knowledge}, pages 184--193, 2015.

\bibitem{lay1998relation}
C.~Lay, A.~Kovacs, and D.~Danto.
\newblock The relation of trait procrastination to the big-five factor conscientiousness: an assessment with primary-junior school children based on self-report scales.
\newblock {\em Personality and Individual Differences}, 25(2):187--193, 1998.

\bibitem{liu2015modeling}
Z.~Liu, J.~He, Y.~Xue, Z.~Huang, M.~Li, and Z.~Du.
\newblock Modeling the learning behaviors of massive open online courses.
\newblock In {\em 2015 IEEE international conference on big data (Big Data)}, pages 2883--2885. IEEE, 2015.

\bibitem{mctighe2005seven}
J.~McTighe and K.~O'Connor.
\newblock Seven practices for effective learning.
\newblock {\em Assessment}, 63(3):10--17, 2005.

\bibitem{midgley2001performance}
C.~Midgley, A.~Kaplan, and M.~Middleton.
\newblock Performance-approach goals: Good for what, for whom, under what circumstances, and at what cost?
\newblock {\em Journal of educational psychology}, 93(1):77, 2001.

\bibitem{mills2016automatic}
C.~Mills, R.~Bixler, X.~Wang, and S.~K. D'Mello.
\newblock Automatic gaze-based detection of mind wandering during narrative film comprehension.
\newblock {\em International Educational Data Mining Society}, 2016.

\bibitem{Nagashima2022}
T.~Nagashima, J.~Britti, X.~Wang, B.~Zheng, V.~Turri, S.~Tseng, and V.~Aleven.
\newblock Designing playful intelligent tutoring software to support engaging and effective algebra learning.
\newblock In {\em Proceedings of the 17th European Conference on Technology-Enhanced Learning (ECTEL 2022)}, volume 13450 of {\em Lecture Notes in Computer Science}, pages 258--271. Springer, 2022.

\bibitem{ocumpaugh2015baker}
J.~Ocumpaugh.
\newblock Baker rodrigo ocumpaugh monitoring protocol (bromp) 2.0 technical and training manual.
\newblock {\em New York, NY and Manila, Philippines: Teachers College, Columbia University and Ateneo Laboratory for the Learning Sciences}, 60, 2015.

\bibitem{panadero2017review}
E.~Panadero.
\newblock A review of self-regulated learning: Six models and four directions for research.
\newblock {\em Frontiers in psychology}, 8:422, 2017.

\bibitem{paquette2014towards}
L.~Paquette, A.~M. de~Carvalho, and R.~S. Baker.
\newblock Towards understanding expert coding of student disengagement in online learning.
\newblock In {\em CogSci}, 2014.

\bibitem{pardos2013affective}
Z.~A. Pardos, R.~S. Baker, M.~O. San~Pedro, S.~M. Gowda, and S.~M. Gowda.
\newblock Affective states and state tests: Investigating how affect throughout the school year predicts end of year learning outcomes.
\newblock In {\em Proceedings of the third international conference on learning analytics and knowledge}, pages 117--124, 2013.

\bibitem{park2018understanding}
J.~Park, R.~Yu, F.~Rodriguez, R.~Baker, P.~Smyth, and M.~Warschauer.
\newblock Understanding student procrastination via mixture models.
\newblock {\em International Educational Data Mining Society}, 2018.

\bibitem{paunonen2001big}
S.~V. Paunonen and M.~C. Ashton.
\newblock Big five predictors of academic achievement.
\newblock {\em Journal of research in personality}, 35(1):78--90, 2001.

\bibitem{pham2015attentivelearner}
P.~Pham and J.~Wang.
\newblock Attentivelearner: improving mobile mooc learning via implicit heart rate tracking.
\newblock In {\em Artificial Intelligence in Education: 17th International Conference, AIED 2015, Madrid, Spain, June 22-26, 2015. Proceedings 17}, pages 367--376. Springer, 2015.

\bibitem{pintrich2004conceptual}
P.~R. Pintrich.
\newblock A conceptual framework for assessing motivation and self-regulated learning in college students.
\newblock {\em Educational psychology review}, 16:385--407, 2004.

\bibitem{raftery1995bayesian}
A.~E. Raftery.
\newblock Bayesian model selection in social research.
\newblock {\em Sociological methodology}, pages 111--163, 1995.

\bibitem{ritter2007cognitive}
S.~Ritter, J.~R. Anderson, K.~R. Koedinger, and A.~Corbett.
\newblock Cognitive tutor: Applied research in mathematics education.
\newblock {\em Psychonomic bulletin \& review}, 14:249--255, 2007.

\bibitem{ritter2016mathia}
S.~Ritter and S.~Fancsali.
\newblock Mathia x: The next generation cognitive tutor.
\newblock In {\em EDM}, pages 624--625. ERIC, 2016.

\bibitem{sabnis2022large}
S.~Sabnis, R.~Yu, and R.~F. Kizilcec.
\newblock Large-scale student data reveal sociodemographic gaps in procrastination behavior.
\newblock In {\em Proceedings of the Ninth ACM Conference on Learning@ Scale}, pages 133--141, 2022.

\bibitem{san2014predicting}
M.~O. San~Pedro, J.~Ocumpaugh, R.~S. Baker, and N.~T. Heffernan.
\newblock Predicting stem and non-stem college major enrollment from middle school interaction with mathematics educational software.
\newblock In {\em EDM}, pages 276--279, 2014.

\bibitem{schunk2012learning}
D.~H. Schunk.
\newblock {\em Learning theories an educational perspective}.
\newblock Pearson Education, Inc, 2012.

\bibitem{shih2011response}
B.~Shih, K.~R. Koedinger, and R.~Scheines.
\newblock A response time model for bottom-out hints as worked examples.
\newblock {\em Handbook of educational data mining}, pages 201--212, 2011.

\bibitem{smallwood2015science}
J.~Smallwood and J.~W. Schooler.
\newblock The science of mind wandering: Empirically navigating the stream of consciousness.
\newblock {\em Annual review of psychology}, 66:487--518, 2015.

\bibitem{stahl2006group}
G.~Stahl.
\newblock {\em Group cognition: Computer support for building collaborative knowledge (acting with technology)}.
\newblock The MIT Press, 2006.

\bibitem{teasley2013constructing}
S.~D. Teasley and J.~Roschelle.
\newblock Constructing a joint problem space: The computer as a tool for sharing knowledge.
\newblock In {\em Computers as cognitive tools}, pages 229--258. Routledge, 2013.

\bibitem{weitnauer2016graspable}
E.~Weitnauer, D.~Landy, and E.~Ottmar.
\newblock Graspable math: Towards dynamic algebra notations that support learners better than paper.
\newblock In {\em 2016 Future Technologies Conference (FTC)}, pages 406--414. IEEE, 2016.

\bibitem{winne2011cognitive}
P.~H. Winne.
\newblock A cognitive and metacognitive analysis of self-regulated learning: Faculty of education, simon fraser university, burnaby, canada.
\newblock In {\em Handbook of self-regulation of learning and performance}, pages 29--46. Routledge, 2011.

\bibitem{winne2017cognition}
P.~H. Winne.
\newblock Cognition and metacognition within self-regulated learning.
\newblock In {\em Handbook of self-regulation of learning and performance}, pages 36--48. Routledge, 2017.

\bibitem{WinneHadwin1998}
P.~H. Winne and A.~F. Hadwin.
\newblock Studying as self-regulated learning.
\newblock In D.~J. Hacker, J.~Dunlosky, and A.~C. Graesser, editors, {\em Metacognition in Educational Theory and Practice}, pages 277--304. Lawrence Erlbaum Associates, Hillsdale, NJ, 1998.

\bibitem{wolters2003regulation}
C.~A. Wolters.
\newblock Regulation of motivation: Evaluating an underemphasized aspect of self-regulated learning.
\newblock {\em Educational psychologist}, 38(4):189--205, 2003.

\bibitem{wolters2021college}
C.~A. Wolters and A.~C. Brady.
\newblock College students’ time management: A self-regulated learning perspective.
\newblock {\em Educational Psychology Review}, 33(4):1319--1351, 2021.

\bibitem{yang2021exploring}
K.~B. Yang, V.~Echeverria, X.~Wang, L.~Lawrence, K.~Holstein, N.~Rummel, and V.~Aleven.
\newblock Exploring policies for dynamically teaming up students through log data simulation.
\newblock {\em International Educational Data Mining Society}, 2021.

\bibitem{yao2020analyzing}
M.~Yao, S.~Sahebi, and R.~Feyzi~Behnagh.
\newblock Analyzing student procrastination in moocs: a multivariate hawkes approach.
\newblock In {\em Proceedings of the 13th Conference on Educational Data Mining (EDM2020)}, 2020.

\bibitem{yonetani2012multi}
R.~Yonetani, H.~Kawashima, and T.~Matsuyama.
\newblock Multi-mode saliency dynamics model for analyzing gaze and attention.
\newblock In {\em Proceedings of the symposium on eye tracking research and applications}, pages 115--122, 2012.

\bibitem{zimmerman1990self}
B.~J. Zimmerman.
\newblock Self-regulated learning and academic achievement: An overview.
\newblock {\em Educational psychologist}, 25(1):3--17, 1990.

\bibitem{zimmerman2002becoming}
B.~J. Zimmerman.
\newblock Becoming a self-regulated learner: An overview.
\newblock {\em Theory into practice}, 41(2):64--70, 2002.

\end{thebibliography}
%
\balancecolumns
\end{document}